\documentclass[a4paper,superscriptaddress,nofootinbib,12pt]{article}
\usepackage{amssymb}
\usepackage{eurosym}
\usepackage{amsfonts}
\usepackage{geometry}
\usepackage{color}
\usepackage{ulem}
\usepackage{bbm}
\usepackage{soul}
\usepackage{color}
\usepackage{mathtools}

\def\be{\begin{equation}}
\def\ee{\end{equation}}
\def\bea{\begin{eqnarray}}
\def\eea{\end{eqnarray}}
\def\h{\hat}
\def\t{\tilde}
\begin{document}
\title{Towards new relativistic doubly $\kappa$-deformed D=4 quantum phase spaces}
\author{J. Lukierski$^a$, S. Meljanac$^b$, S. Mignemi$^c$, A. Pacho\l$^d$ and M. Woronowicz$^a$}
\date{}
\maketitle
\mbox{}\\
\footnotesize{$^a$ Institute of Theoretical Physics, Wroc\l aw University, pl.~Maxa Borna 9, 50-205 Wroc\l aw, Poland\\
$^b$ Division of Theoretical Physics, Rudj{}er
Bo\v{s}kovi\'c Institute, Bijeni\v{c}ka~c.54, HR-10002~Zagreb, Croatia\\
$^c$ Dipartimento di Matematica, Universit\`a di
Cagliari via Ospedale 72, 09124 Cagliari, Italy and INFN, Sezione di
Cagliari, 09042 Monserrato, Italy\\
$^d$ Department of Microsystems, University of South-Eastern Norway, Campus Vestfold, Norway}
\begin{abstract}
We propose new noncommutative models of quantum phase spaces, containing a pair of $\kappa$-deformed Poincar\'e algebras, with two independent double ($\kappa,\tilde{\kappa}$)-deformations in  space-time and four-momenta sectors.
The first such quantum phase space can be obtained by
contractions $M,R\to \infty$ of recently introduced doubly $\kappa$-deformed $(\kappa,\tilde{\kappa})$-Yang models, with the parameters $M,R$ describing inverse space-time and four-momenta curvatures and constant four-vectors $a_\mu, b_\mu$ determining nine types of $(\kappa,\tilde{\kappa})$-deformations. The second considered model is provided by the nonlinear doubly $\kappa$-deformed TSR algebra spanned by 14 coset $\hat{o}(1,5)/\hat {o}(2)$ generators. The
basic algebraic difference between the two models is the following: the first one, described by $\hat{o}(1,5)$ Lie algebra can be supplemented by the Hopf algebra structure, while the second model contains the quantum phase space commutators $[\hat{x}_\mu,\hat{q}_\nu]$, with the standard numerical  $i\hbar\eta_{\mu\nu}$ term; therefore it describes the quantum-deformed Heisenberg algebra relations which cannot be equipped with the Hopf algebra. 
\end{abstract}

\section{Introduction}

 The principles of quantum theory and general relativity - the two pillars of modern physics - are incompatible because they describe two different areas of physical reality. While quantum theory is the theory of microscopic scales, the general relativity governs the macroscopic realm, describing nature at medium and large distances. Reconciling these different frameworks requires new concepts on how to alter both theories and  neither quantum mechanics nor general relativity can be regarded as sufficient and definitive.  
To explore deviations from quantum theory  and to allow for embedding of the gravitational effects in the quantum scale, one should study the new models of quantum mechanical phase spaces introducing quite radical
modifications to the canonical Heisenberg relations. However, any changes to the Heisenberg relations between the observables of position and momenta will imply modifications of phase space structure and of respective Uncertainty Principle (UP). Various generalizations of UP have been considered in the literature and were related with new models of quantum phase spaces, see e.g. \cite{Maggiore:1993rv}-
\cite{MM_2411}.

Probing the modifications arising in quantum mechanical phase spaces has been gaining interest from an experimental point of view, see e.g. \cite{Nature}-
\cite{Grav}. However, the bounds on the modification (deformation) parameters (which are linked with the quantum gravity (QG) corrections) that are obtained from experiments are still far away from the desired values. The recent summary of bounds obtained from the table-top and astrophysical experiments can be found in Ref. \cite{Bosso}.
To significantly enhance our ability to probe nature in regimes where gravity intersects with quantum physics, advancements in the table-top (or astrophysical) experiments are needed. Nevertheless, the precision required to explore the interplay between gravity and quantum systems at ultra-low energies may soon be achieved, hence studying different models of deformed quantum phase spaces is very timely. The more precise experiments focusing on interaction between gravity and quantum physics may shed some light on finding quantum gravity effects, potentially revealing the first signatures of a new physics.

In this work we propose the new relativistic deformed quantum phase spaces which contain two mass-like deformation parameters $\kappa$ and $\tilde{\kappa}$ together with the parameters describing the curvatures of noncommutative (NC) space-time as well as the NC momentum space. Recently \cite{PLB2024} (see also \cite{MB-MM}, \cite{PoS23}), the double $\kappa$-deformation of the Yang model\footnote{\tiny{In short-cut notation it is also called $(\kappa,\tilde{\kappa})$-Yang model, see \cite{PLB2024}.}} has been introduced, containing as its subalgebras the $\kappa$-deformation of NC coordinates and the $\tilde{\kappa}$-deformation of NC fourmomentum space. Therein, it was shown how to obtain the doubly $\kappa$-deformed Yang models from the $\kappa$-deformed Snyder models \cite{kappa-Snyder1}-\cite{21} by applying the generalized Born map. \\

In this paper we propose new models of relativistic doubly $\kappa$-deformed D=4 quantum phase spaces which can be obtained in two different ways.\\
The first way, which we present in Sec. 3, is to derive the new models from the doubly $\kappa$-deformed Yang models \cite{PLB2024,MB-MM,PoS23}. Within this approach we consider the following two alternative derivations:\\
i) Firstly, we perform the contraction procedure $(M,R)\to\infty$ in the doubly $\kappa$-deformed Yang model and obtain new relativistic doubly $\kappa$-deformed D=4 quantum phase space, depending on two parameters $\kappa$ and $\tilde{\kappa}$.\\
ii) Secondly, we use the Lie algebraic description of doubly $\kappa$-deformed Yang model as represented by $\hat o(1,5;g)$ Lie algebra which can be obtained from $\hat{o}(1,5)\equiv\hat{o}(1,5,\eta)$ by replacing diagonal pseudoorthogonal metric $\eta_{AB}$ by suitably chosen constant parameter-dependent symmetric metric $g\equiv g_{AB}$\footnote{\tiny{In fact we choose the algebras $\hat o(1,5;g)$ which are isomorphic to the classical Lie algebra $\hat o(1,5)$}}. We properly fix the parameter-dependent metric $g$ to obtain the new model of relativistic doubly $\kappa$-deformed D=4 quantum phase space, depending on three independent parameters $\kappa,\tilde{\kappa}$ and $\rho$.\\
In both cases i) and ii) the obtained doubly $\kappa$-deformed quantum phase spaces are algebraically equivalent to $\hat{o}(1,5)$ Lie algebra. i.e. to the original Yang model, which is recalled in Sec. 2.\\

Second way to obtain the new model of relativistic doubly $\kappa$-deformed $D=4$ quantum phase space is to consider
another possible algebra with NC coordinates and NC four-momenta which is provided by the Triply
Special Relativity (TSR) model\footnote{\tiny{Often, TSR model is also called Snyder-de Sitter (SdS) model since it is obtained as a generalization of the Snyder model (flat space-time with curved momentum space) to the model with curved both NC space-time and NC four-momenta.}} \cite{TSR}. It is spanned by 14 generators of the coset $\hat{o}(1,5)/\hat{o}(2)$ and depends on three independent fundamental parameters. 
Additionally, in TSR model, one considers appropriate deformations of the Heisenberg phase space commutator $[\hat x_\mu,\hat q_\nu]$, which in the quantum-mechanical limit, without QG corrections, reduces to the canonical Heisenberg commutation relations.

The main novelty of the present paper is in Sec. 4, where we propose the modification of TSR model
by introducing doubly $\kappa$-deformation in the quantum phase space. In such a case the $(\kappa,\tilde{\kappa})$-deformed quantum phase space reduces to the canonical Heisenberg relations in the contraction limit $\kappa\to\infty$ and $\tilde{\kappa}\to 0$.

\section{Yang Models - Preliminary Considerations}
 To introduce the new models of relativistic quantum phase spaces, we start with the canonical $D=4$ relativistic phase space algebra as generated by commuting space-time  $x_\mu$ and four-momenta $q_\mu$ coordinates and can be extended by the Lorentz symmetry generators  $\hat{M}_{\mu\nu}$. We denote such algebra by $H^{(1,3)}=(x_\mu,q_\mu,\hat{M}_{\mu\nu})$, with space-time coordinates and four-momenta satisfying the Heisenberg canonical commutation relations:
\begin{equation}\label{xqQM}
[x_\mu,q_\nu]=i\hbar\eta_{\mu\nu}
\end{equation}
where $\mu,\nu=0,1,2,3$, $\eta_{\mu\nu}=\mathrm{diag}(-1,1,1,1)$ and Lorentz algebra
\begin{equation}
\lbrack \hat{M}_{\mu \nu },\hat{M}_{\rho \tau }]=i\hbar (\eta _{\mu \rho }%
\hat{M}_{\nu \tau }-\eta _{\mu \tau }\hat{M}_{\nu \rho }+\eta _{\nu \tau }%
\hat{M}_{\mu \rho }-\eta _{\nu \rho }\hat{M}_{\mu \tau }) \label{MM}
\end{equation} with the defining relativistic covariance relations:
\begin{equation}
\lbrack \hat{M}_{\mu \nu },{x}_{\rho }]=i\hbar (\eta _{\mu \rho }{x}%
_{\nu }-\eta _{\nu \rho }{x}_{\mu }),  \qquad
\lbrack \hat{M}_{\mu \nu },{q}_{\rho }]=i\hbar (\eta _{\mu \rho }{q}%
_{\nu }-\eta _{\nu \rho }{q}_{\mu }). \label{Mqcl}
\end{equation}
From the set of the generators of algebra $H^{(1,3)}$ one can construct the pair of overlapping classical Poincar\'e algebras: one generated by $(x_\mu,\hat{M}_{\mu\nu})$
and the second one by $(q_\mu,\hat{M}_{\mu\nu})$.
The relations \eqref{xqQM}-\eqref{Mqcl} are covariant under the following Born map $B$ \cite{Born1,Born2,Freidel}
\begin{equation}
B:\quad {x}_{\mu }\rightarrow {q}_{\mu },\quad {q}_{\mu
}\rightarrow -{x}_{\mu },\quad\hat{M}_{\mu \nu
}\leftrightarrow \hat{M}_{\mu \nu }
\label{born}
\end{equation}
and it follows from relations \eqref{xqQM}-\eqref{Mqcl} that the algebra $H^{(1,3)}$ is Born-selfdual.

Our first aim will be to describe the deformations of the algebra $H^{(1,3)}$ which lead to the noncanonical algebra $\hat{H}^{(1,3)}$ with NC quantum phase space coordinates $(\hat{x}_\mu,\hat{q}_\mu)$. Due to the presence of QG effects (see e.g. \cite{DFR}) we replace $x_\mu\rightarrow \hat x_\mu,\quad q_\mu\rightarrow \hat q_\mu$, and after the quantization, we get from $H^{(1,3)}$ the quantum relativistic algebra $H^{(1,3)}\rightarrow\hat H^{(1,3)}=(\hat x_\mu,\hat q_\mu,\hat{M}_{\mu\nu})$. Then in general $[\hat x_\mu,\hat x_\nu]\neq 0$ and $[\hat q_\mu, \hat q_\mu]\neq 0$, while the commutator $[\hat x_\mu,\hat q_\nu]$ becomes significantly different from the relation \eqref{xqQM}.

In the following we will consider two well-known NC models which provide the particular choices of the algebras $\hat H^{(1,3)}$. The first example of such algebra has been proposed by Yang \cite{Yang} as an extension of the Snyder model \cite{Snyder}. The defining relations of $D=4$ Yang model are as follows:
\begin{equation}
\lbrack \hat{x}_{\mu },\hat{x}_{\nu }]=\frac{i\hbar }{M^{2}}\hat{M}_{\mu \nu
},  \qquad
\lbrack \hat{q}_{\mu },\hat{q}_{\nu }]={\frac{i\hbar }{R^{2}}}\hat{M}_{\mu
\nu },\qquad
\lbrack \hat{x}_{\mu },\hat{q}_{\nu }]=i\hbar \eta _{\mu \nu }\hat{r}
\label{xq}
\end{equation}
where the (real) parameters $M$ and $R$ describe the curvatures of NC space-time and NC
four-momenta space\footnote{\tiny{For the notions and the use of NC Riemannian geometry see e.g. \cite{5a,5b,5c,5d} in quantum gravity models and for example string theory \cite{5e}}}.
The additional generator $\hat r$ satisfies the relations:
\begin{equation}
\lbrack \hat{r},\hat{x}_{\mu }]=\frac{i\hbar }{M^{2}}\hat{q}_{\mu },\qquad
\lbrack \hat{r},\hat{q}_{\mu }]=-\frac{i\hbar }{R^{2}}\hat{x}_{\mu }
\label{rxrp}
\end{equation}
and it follows that $\hat r$ describes the generator of particular $\hat{o}(2)$ internal symmetries \footnote{\tiny{The standard $\hat{o}(2)$ relations with $\hat{o}(2)$ generator $\hat I$ are $\lbrack I,\hat{X}_{\mu }]=i\hbar \hat{Q}_{\mu },\quad
\lbrack I,\hat{Q}_{\mu }]=- i\hbar \hat{X}_{\mu }$ and can be obtained from \eqref{rxrp} by the redefinition: $\hat I=MR \hat{r},\quad \hat{X}_\mu=M\hat{x}_\mu,\quad \hat{Q}_\mu=R\hat{q}_\mu$.
}}.

If we supplement the Born map \eqref{born} with the following additional mappings:
\begin{equation}\label{MRrr}
B:\qquad
M \leftrightarrow R,\qquad \hat{r}\leftrightarrow\hat{r}
\end{equation}
it is easy to see that the Yang model \eqref{xq}-\eqref{rxrp} is Born self-dual. After the following assignment of the generators:
\begin{equation}
\hat{M}_{\mu 4 }=M\hat{x}_{\mu },\
\hat{M}_{\mu 5}=R\hat{q}_{\mu },\ \hat{M}_{45}=MR\hat{r}
\label{M_Y_AB}
\end{equation}
the relations \eqref{xq}-\eqref{rxrp} and \eqref{MM} permit to interpret the $D=4$ Yang algebra as algebraically equivalent to $\hat{o}(1,5)$ Lie algebra:
\begin{equation}
\lbrack \hat{M}_{AB},\hat{M}_{CD}]=i\hbar (\eta _{AC}\hat{M}_{BD}-\eta _{AD}
\hat{M}_{BC}+\eta _{BD}\hat{M}_{AC}-\eta _{BC}\hat{M}_{AD})  \label{YangMM}
\end{equation}%
where $\eta _{AB}=diag(-1,1,\ldots ,1)$ and $A,B=0,1,\ldots ,5$.

Another example of NC models which will be considered here is provided by the NC geometry of $\kappa$-deformed Minkowski space-time accompanied by Born-dual $\tilde{\kappa}$-deformed four-momenta space. In order to describe all three possible types of $\kappa$-deformations of NC Minkowski space-time (time-like, standard or light-like) one introduces the constant fourvector $a_\mu$ which selects the quantized direction in space-time \cite{kMvec1}
-\cite{kMvec4}. The $\kappa$-Minkowski commutation relations are \cite{kappa-Mink1,kappa-Mink2}:
\begin{equation}\label{xxk}
\lbrack \hat{x}_\mu,\hat{x}_{\nu }]=\frac{i\hbar }{\kappa}({a}_{\mu }\hat{x}_\nu-a_\nu \hat{x}_\mu)
\end{equation}
where $a_\mu$ can be chosen in threefold way as $a^2=a_\mu a^\mu=(1,0,-1)$\footnote{\tiny{The relations \eqref{xxk},\eqref{qqk} introduce two mass-like parameters $\kappa$, $\tilde{\kappa}$ and two constant four-vectors $a_\mu$, $b_\mu$ limited by threefold constraints $a^2=(1,0,-1)$ and $b^2=(1,0,-1)$ which specify $3\times 3=9$ types of $(\kappa,\tilde{\kappa})$-deformed pairs of Poincar\'e algebras. In general cases, as we will show in Sec. 3, still there should appear a third parameter $\rho$.}}.
 By introducing the constant fourvector $b_\mu$ describing quantization direction in quantum fourmomentum space $\hat{q}_\mu$
one can introduce the following analogous $\tilde{\kappa}$-deformation of NC four-momenta\footnote{\tiny{In general case the mass-like parameters $\kappa$ and $\tilde{\kappa}$ are independent.}}:
\begin{equation}  \label{qqk}
\lbrack \hat{q}_{\mu },\hat{q}_{\nu }]=i\hbar \tilde{\kappa}(b_{\mu }\hat{q}%
_{\nu }-b_{\nu }\hat{q}_{\mu }).
\end{equation}

\section{Doubly $\kappa$-deformed D=4 quantum phase spaces from Yang models}
Many efforts to generalize Snyder and Yang models appeared in the
literature, for example the $\kappa$-Minkowski extension of Snyder model was firstly proposed in \cite{kappa-Snyder1}
(see also \cite{kappa-Snyder2}-\cite{Lukierski:2022cpx}) while $\kappa$-Minkowski type extensions of Yang model
in both coordinates and four-momenta sectors were proposed recently in \cite{PLB2024} (see also \cite{MB-MM}, \cite{PoS23}). In
such extensions of the Yang model we need to introduce a pair of different $
\kappa $-Minkowski terms, respectively in NC space-time and quantum
four-momenta, i.e. we need two independent mass-like parameters $%
\kappa $ and $\tilde{\kappa}$ resulting in $(\kappa ,\tilde{\kappa%
})$-Yang model, which can be defined by the following set of commutation
relations (see \cite{PLB2024}):
\begin{equation}
\lbrack \hat{x}_{\mu },\hat{x}_{\nu }]=i\hbar \left[ \frac{1}{M^{2}}\hat{M}%
_{\mu \nu }+\frac{1}{\kappa }(a_{\mu }\hat{x}_{\nu }-a_{\nu }\hat{x}_{\mu })%
\right] ,  \qquad
\lbrack \hat{q}_{\mu },\hat{q}_{\nu }]=i\hbar \left[ \frac{\hat{M}_{\mu \nu }%
}{R^{2}}+\tilde{\kappa}(b_{\mu }\hat{q}_{\nu }-b_{\nu }\hat{q}_{\mu })\right]
\label{qqYk}
\end{equation}%
and
\begin{equation}
\lbrack \hat{M}_{\mu \nu },\hat{x}_{\rho }]=i\hbar \left[ \eta _{\mu \rho }%
\hat{x}_{\nu }-\eta _{\nu \rho }\hat{x}_{\mu }+\frac{1}{\kappa }(a_{\mu }%
\hat{M}_{\rho \nu }-a_{\nu }\hat{M}_{\rho \mu })\right] ,  \label{MxYk}
\end{equation}%
\begin{equation}
\lbrack \hat{M}_{\mu \nu },\hat{q}_{\rho }]=i\hbar \left[ \eta _{\mu \rho }%
\hat{q}_{\nu }-\eta _{\nu \rho }\hat{q}_{\mu }+\tilde{\kappa}(b_{\mu }\hat{M}%
_{\rho \nu }-b_{\nu }\hat{M_{\rho \mu }})\right] .  \label{MqYk}
\end{equation}
We see that in the relations \eqref{qqYk} the noncommutativity of quantum space-time and four-momenta is described by the sum of the noncommutativties from the Yang model and from $\kappa$-deformation.
Further, it can be shown that the relations \eqref{MxYk}, \eqref{MqYk} can be justified by the calculation of Jacobi identities. The remaining  quantum phase space relations are \cite{PLB2024}, \cite{MB-MM} \footnote{\tiny{To link the notations used in \cite{PLB2024} and \cite{MB-MM} we need the following convention matching, where the left hand side contains the notation used in
\cite{PLB2024} and the right hand side corresponds to the one from \cite{MB-MM}:
\[
\mu ,\nu \rightarrow i,j;\quad \hbar =1;\quad\frac{1}{R}=\alpha \quad \frac{1%
}{M}=\beta;
\]
and the generators
\[
\hat{x}_{\nu }\rightarrow \tilde{X}_{i};\quad\hat{q}_{\mu }\rightarrow
\tilde{P}_{i};\quad\hat{r}\rightarrow \hat{h};\quad{a}_{\mu }\to\kappa\beta a_{\mu
};\quad {b}_{\mu }\rightarrow (\alpha/\tilde{\kappa})b_{\mu }.
\]%
}}:
\begin{equation}
\lbrack \hat{x}_{\mu },\hat{q}_{\nu }]=i\hbar \left( \eta _{\mu \nu }\hat{r}+%
\tilde{\kappa}b_{\mu }\hat{x}_{\nu }-\frac{a_{\nu }}{\kappa }\hat{q}_{\mu }+%
\frac{\rho }{MR}\hat{M}_{\mu \nu }\right) , \label{xqYk}
\end{equation}%
\begin{equation}
\lbrack \hat{r},\hat{x}_{\mu }]=i\hbar \left( \frac{1}{M^{2}}\hat{q}_{\mu }-%
\frac{1}{MR}\rho \hat{x}_{\mu }-\frac{a_{\mu }}{\kappa }\hat{r}\right) , \qquad
\lbrack \hat{r},\hat{q}_{\mu }]=i\hbar \left( -\frac{1}{R^{2}}\hat{x}_{\mu }+%
\frac{1}{MR}\rho \hat{q}_{\mu }-\tilde{\kappa}b_{\mu }\hat{r}\right) .
\label{rqYk}
\end{equation}
Additionally, we have
\begin{equation}
\lbrack \hat{r},\hat{M}_{\mu \nu }]=-i\hbar \left[ \frac{1}{\kappa }(a_{\mu }%
\hat{q}_{\nu }-a_{\nu }\hat{q}_{\mu })-\tilde{\kappa}(b_{\mu }\hat{x}_{\nu
}-b_{\nu }\hat{x}_{\mu })\right]. \label{rmYk}
\end{equation}
Relation \eqref{rmYk} shows that the internal and Lorentzian generators do not commute with
each other.
The $\kappa,\tilde{\kappa} $-dependence of the Yang model can be explained by the
generalized Born map $\tilde{B}$, obtained if we supplement \eqref{born} and \eqref{MRrr} by the following relations\footnote{\tiny{In standard Born duality  (e.g. for Yang algebra (\ref{xq})-(\ref{rxrp})) the algebra is self-dual due to the exchange of generators and constants responsible for agreement of dimensions of the generators (see (\ref{born}) and (\ref{MRrr})).
The double $\kappa$ deformed Yang algebra (\ref{qqYk})-(\ref{rmYk}) is self-dual in the above sense only in the case when  the four-vectors $a=b=0$. For other values of $a$ and $b$ the double $\kappa$ deformed Yang algebra is self-dual only if we additionally require the change of the structure constants of algebra (corresponding to $a\rightarrow b, b\rightarrow -a$) and we name such an extension of Born duality as generalized Born duality.
}}
\begin{equation}
\tilde{B}:\quad a_{\mu }\rightarrow b_{\mu },\quad b_{\mu }\rightarrow
-a_{\mu },\quad \kappa \leftrightarrow \frac{1}{\tilde{\kappa}}.
\label{born2}
\end{equation}
Both maps \eqref{born}, \eqref{born2} describe pseudo-involutions, satisfying the relations $B^2=\tilde{B}^4=1$.
It can be checked that the relations (\ref{xqYk})-(\ref{rmYk}) are self dual
under the generalized Born map (\ref{born}), extended by (\ref{born2}).

\subsection{Doubly $\kappa$-deformed D=4 quantum phase spaces by contraction of $(\kappa,\tilde{\kappa})$-Yang model}
To obtain the new model of quantum phase space, including both NC coordinates and NC four-momenta, and which is
dependent on two additional parameters
$\kappa$ and $\tilde{\kappa}$, we perform the contraction when $M\to\infty $ and $R\to\infty$ in doubly $\left(\kappa ,\tilde{\kappa}\right)$-Yang model, described by \eqref{qqYk}-\eqref{rmYk}. We call this new model doubly $\kappa$-Poincar\'e algebra.
It is given by relations\footnote{\tiny{Relations \eqref{xxk}, \eqref{qqk} are repeated here for convenience and clarity of presentation of the new model.}}
\begin{equation}\label{xxk1}
\lbrack \hat{x}_\mu,\hat{x}_{\nu }]=\frac{i\hbar }{\kappa}({a}_{\mu }\hat{x}_\nu-a_\nu \hat{x}_\mu), \qquad
\lbrack \hat{q}_{\mu },\hat{q}_{\nu }]=i\hbar \tilde{\kappa}(b_{\mu }\hat{q}%
_{\nu }-b_{\nu }\hat{q}_{\mu }),
\end{equation}
\begin{equation}  \label{Mxk}
\lbrack \hat{M}_{\mu \nu },\hat{x}_{\rho }]=i\hbar \left[ \eta _{\mu \rho }%
\hat{x}_{\nu }-\eta _{\nu \rho }\hat{x}_{\mu }+\frac{1}{\kappa }(a_{\mu }%
\hat{M}_{\rho \nu }-a_{\nu }\hat{M}_{\rho \mu })\right] ,
\end{equation}%
\begin{equation}  \label{Mqk}
\lbrack \hat{M}_{\mu \nu },\hat{q}_{\rho }]=i\hbar \left[ \eta _{\mu \rho }%
\hat{q}_{\nu }-\eta _{\nu \rho }\hat{q}_{\mu }+\tilde{\kappa}(b_{\mu }\hat{M}%
_{\rho \nu }-b_{\nu }\hat{M_{\rho \mu }})\right],
\end{equation}
\begin{equation}  \label{xqk}
\lbrack \hat{x}_{\mu },\hat{q}_{\nu }]=i\hbar \left( \eta _{\mu \nu }\hat{r}+%
\tilde{\kappa}b_{\mu }\hat{x}_{\nu }-\frac{a_{\nu }}{\kappa }\hat{q}_{\mu
}\right) ,
\end{equation}
\begin{equation}  \label{rxk}
\lbrack \hat{r},\hat{x}_{\mu }]=-i\hbar \frac{a_{\mu }}{\kappa }\hat{r},\qquad  
\lbrack \hat{r},\hat{q}_{\mu }]=-i\hbar \tilde{\kappa}b_{\mu }\hat{r},
\end{equation}
\begin{equation}  \label{rmk}
\lbrack \hat{r},\hat{M}_{\mu \nu }]=-i\hbar \left[ \frac{1}{\kappa }(a_{\mu }%
\hat{q}_{\nu }-a_{\nu }\hat{q}_{\mu })-\tilde{\kappa}(b_{\mu }\hat{x}_{\nu
}-b_{\nu }\hat{x}_{\mu })\right].
\end{equation}
\subsection{Doubly $\kappa$-deformed $D=4$ quantum phase spaces by fixing the metric $g_{AB}$ in  $\hat o(1,5, g_{AB})$}
It is already known since 1947 (see \cite{Yang}) that the algebra describing
$D=4$ Yang model, i.e. \eqref{xq}-\eqref{rxrp} is spanned by the $\hat{o}(1,5)$ algebra. The generalizations of this description, by replacing $\eta _{AB}$ by more general metric $g_{AB}$ ($\eta_{AB}\to g_{AB}$) have been proposed, e.g. in \cite{AB-AP-EPJC}. Doubly $(\kappa ,\tilde{\kappa})$-Yang models, which we recalled in the beginning of Sec.\ 3 and which are described by eqs \eqref{qqYk}-\eqref{rmYk}, have the Lie algebra structure of $\hat{o}(1,5;g^{(Y)}_{AB})$ (see \cite{PLB2024}):
\begin{equation}
\lbrack \hat{M}_{AB},\hat{M}_{CD}]=i\hbar (g_{AC}^{(Y)}\hat{M}%
_{BD}-g_{AD}^{(Y)}\hat{M}_{BC}+g_{BD}^{(Y)}\hat{M}_{AC}-g_{BC}^{(Y)}\hat{M}%
_{AD})  \label{Yang_gMM}
\end{equation}%
where the symmetric metric components $g_{AB}^{(Y)}$ with the signature $
(-1,1,\ldots ,1)$ depend on five parameters
\footnote{\tiny{Note that in Sec.\ 3.2 the parameters $M,R$ are kept finite.}}
$(M,R,\kappa ,\tilde{\kappa},\rho )$ 
where $(M=\lambda ^{-1},\tilde{M}=R^{-1})$ is the pair of mass parameters (or equivalently the pair of length parameters $\lambda =M^{-1},R$), the
mass-like parameters $\left( \kappa ,\tilde{\kappa}\right) $ and the
dimensionless parameter $\rho $
and a pair of constant dimensionless four-vectors $a_{\mu
}, b_{\mu }$, ($\mu =0,1,2,3$) which respectively determine
the type of $\kappa $-dependence in $D=4$ quantum space-time and $D=4$
quantum four-momenta sectors of Yang algebra. The metric $g_{AB}^{(Y)}$ is
determined by the following assignments of the generators:
\begin{equation}
\hat M_{AB}=\left( \hat{M}_{\mu \nu },\ \hat{M}_{\mu 4 }=M\hat{x}_{\mu },\
\hat{M}_{\mu 5 }=R\hat{q}_{\mu },\ \hat{M}_{45}=MR\hat{r}\ \right)
\label{M_Y_AB}
\end{equation}%
where $[M_{AB}]=L^{0}$ (dimensionless), in consistency with relation (\ref%
{Yang_gMM}), with $\hat{M}_{\mu \nu }$ describing $D=4$ Lorentz algebra and
the scalar $\hat{r}$ providing the generator of the $\hat{o}(2)$ internal
symmetries. Relations (\ref{Yang_gMM}) describe the $(\kappa ,\tilde{%
\kappa})$-Yang model if we insert the following components of the $D=6$
metric tensor:
\begin{equation}
g_{AB}^{\left( Y\right) }=\left(
\begin{array}{ccc}
\eta _{\mu \nu } & g_{\mu 4}^{\left( Y\right) } & g_{\mu 5}^{\left( Y\right)
} \\
g_{4\nu }^{\left( Y\right) } & g_{44}^{\left( Y\right) } & g_{45}^{\left(
Y\right) } \\
g_{5\nu }^{\left( Y\right) } & g_{54}^{\left( Y\right) } & g_{55}^{\left(
Y\right) }%
\end{array}%
\right)  \label{g_matrix}
\end{equation}%
where\footnote{\tiny{We note that the metric $g_{AB}$ with fixed $\eta_{\mu\nu}$ can in the most general case be built up from 11 free parameters with arbitrary values of $g_{44}$ and $g_{55}$ (see also \cite{MB-MM}). However, following our choice in \cite{PLB2024} we are using the version with only 9 free parameters, fixing 2 parameters by the relations $g_{44} = 1$ and $g_{55} = 1$.}}
\begin{equation}
g_{\mu 4}^{\left( Y\right) }=g_{4\mu }^{\left( Y\right) }=\frac{M}{\kappa }%
a_{\mu },\,\quad g_{\mu 5}^{\left( Y\right) }=g_{5\mu }^{\left( Y\right) }=R%
\tilde{\kappa}b_{\mu },\quad g_{45}^{\left( Y\right) }=g_{54}^{\left(
Y\right) }=\rho ,\quad g_{44}^{\left( Y\right) }=g_{55}^{\left( Y\right) }=1.
\label{g_coeff}
\end{equation}
Note $g_{AB}^{(Y)}$ are dimensionless ($[g_{AB}^{(Y)}]=L^{0}$) in consistency with relations (\ref{Yang_gMM}). 

To obtain the new model of doubly $\kappa$-deformed quantum phase space (doubly $\kappa$-Poincar\'e algebra), depending on three parameters $\kappa$, $\tilde{\kappa}$ and $\rho$ we modify the metric $g^{(Y)}_{AB}$ by setting $g_{44}=0=g_{55}=0$. We note that this new quantum phase space will preserve Lorentz covariance and we consider the same assignment of
generators (\ref{M_Y_AB}) but only change the metric $g_{AB}^{\left( Y\right) }
$ to the following
\begin{equation}\label{metric}
\tilde g_{AB}=\left(
\begin{array}{ccc}
\eta _{\mu \nu } & g_{\mu 4}=\frac{M}{\kappa }a_{\mu } & g_{\mu 5}=R\tilde{%
\kappa}b_{\mu } \\
g_{4\nu }=\frac{M}{\kappa }a^T_{\nu } & 0 & \rho \\
g_{5\nu }=R\tilde{\kappa}b^T_{\nu } & \rho & 0%
\end{array}%
\right)
\end{equation}
where we require
\begin{equation}\label{detg}
\det\tilde{g}_{AB}=\rho ^{2}-2\rho MR\frac{\tilde{\kappa}}{\kappa }\left(a_{\mu }b^{\mu }\right) +\left( MR\frac{\tilde{\kappa}}{\kappa }\right)
^{2}[\bigskip \left( a_{\mu }b^{\mu }\right) ^{2}-a^{2}b^{2}]\neq 0.
\end{equation} Now the relations of $\hat{o}(1,5,\tilde g_{AB})$:
\begin{equation}
\lbrack \hat{M}_{AB},\hat{M}_{CD}]=i\hbar (\tilde g_{AC}\hat{M}_{BD}-\tilde g_{AD}\hat{M}%
_{BC}+\tilde g_{BD}\hat{M}_{AC}-\tilde g_{BC}\hat{M}_{AD})
\end{equation}
describe doubly $\kappa$-deformed quantum phase space algebra given by \eqref{xxk1}, \eqref{Mxk}, \eqref{Mqk} and \eqref{MM} with the exception of the commutation relation between NC coordinates and NC four-momenta which now becomes:
\begin{equation}\label{xq--}
\lbrack \hat{x}_\mu,\hat{q}_\nu ]=
i\hbar (\eta _{\mu \nu }\hat{r}+\tilde{\kappa}b_{\mu }\hat{x}_{\mu }-\frac{a_{\nu }}{\kappa }\hat{q}_{\mu }+\frac{\rho}{MR} \hat{M}_{\mu \nu })
\end{equation}

Note that \eqref{xq--} provides a more general  version of the doubly $\kappa$-deformed phase space than the one obtained in the previous section, cf. \eqref{xqk}.
Because of the last term, proportional to the parameter $\rho$, the two models of relativistic $D=4$ doubly $\kappa$-deformed quantum phase spaces, called doubly $\kappa$-Poincar\'e algebras (obtained in Sec.\ 3.1 and 3.2) are different. 

We can obtain from the relation \eqref{xq--}, the version from Sec.\ 3.1 by setting $\rho = 0$ in \eqref{metric}, while keeping $\det\tilde{g}_{AB}\neq 0$.
These conditions put certain restrictions on the possible choices of the constant four-vectors $a_\mu$ and $b_\mu$, namely 
$(a_\mu b^\mu)^2\neq a^2 b^2$. One of the allowed cases would be $a_\mu b^\mu = 0$, $a^2\neq 0$, $b^2\neq 0$ (see \eqref{detg}). In particular these restrictions  imply that $b^\mu$ cannot be proportional to $a^\mu$.

\subsection{Doubly $\kappa$-deformation directly from Yang model}
In Sections 3.1 and 3.2 we performed the reductions of doubly $(\kappa ,\tilde{\kappa})$-Yang models \cite{PLB2024} \eqref{qqYk}-\eqref{rmYk} to obtain double-$\kappa$ deformation. In this Section we present a more straightforward possibility, by starting from the Yang algebra \eqref{MM},  \eqref{xq}-\eqref{rxrp} and  introducing the change of basis resulting in the double-$\kappa$ deformation. To obtain the full set of Yang algebra commutators we add to \eqref{MM}, \eqref{xq}-\eqref{rxrp} the relation
$[\h{r},\h{M}_{\mu\nu}]=0 $
what implies that $\hat{r}$ describes an Abelian internal $\hat{o}(2)$ symmetry generator.
Then, we can change the basis of \eqref{MM},  \eqref{xq}-\eqref{rxrp}  by the following transformation (see e.g.\cite{baza})
\begin{equation}
\tilde{x}_\mu=\hat{x}_\mu+\frac{1}{\kappa} a^\rho M_{\mu\rho},\qquad \tilde{q}_\mu=\hat{q}_\mu+\t{\kappa} b^\rho M_{\mu\rho},\qquad \tilde{M}_{\mu\nu}=\hat{M}_{\mu\nu}\label{YT}
\end{equation}
where $\kappa$, $\t{\kappa}$ are real parameters and $a_\mu$, $b_\mu$ real four-vectors as before. At this stage we do not specify the dependence of the $\h{r}$ generator on the remaining Yang algebra generators
$\t{x}_\mu,\t{q}_\mu,\t{M}_{\mu\nu}$ generators. The algebra of the transformed generators \eqref{YT} looks as follows:
\begin{equation}
[\tilde{x}_\mu,\tilde{x}_\nu]=i\hbar(\frac{1}{M^2}+\frac{a^2}{\kappa^2})\tilde{M}_{\mu\nu}+\frac{i\hbar}{\kappa}(a_\mu\tilde{x}_\nu-a_\nu\tilde{x}_\mu),\label{Ygxx}
\end{equation}
and
\begin{equation}
[\tilde{q}_\mu,\tilde{q}_\nu]=i\hbar(\frac{1}{R^2}+\t{\kappa}^2 b^2)\tilde{M}_{\mu\nu}+i\hbar\t{\kappa}(b_\mu\tilde{q}_\nu-b_\nu\tilde{q}_\mu),\label{Ygpp}
\end{equation}
where $a^2=a^\mu a_\mu$ and $b^2=b^\mu b_\mu$ and we additionally have:
\begin{equation}
[\t{x}_\mu,\t{q}_\nu]=i\hbar\eta_{\mu\nu}(\h{r}+\frac{1}{\kappa} a^\rho\t{q}_\rho-\t{\kappa} b^\rho \t{x}_\rho-\frac{\t{\kappa}}{\kappa} a^\rho b^\sigma \t{M}_{\rho\sigma})+i\hbar\left[\t{\kappa} b_\mu\t{x}_\nu-\frac{1}{\kappa} a_\nu\t{q}_\mu+\frac{\t{\kappa}}{\kappa} a^\rho b_\rho \t{M}_{\mu\nu}\right],\label{Yxp}
\end{equation}
\begin{equation}
[\tilde{M}_{\mu\nu},\tilde{x}_\rho]=i\hbar (\eta _{\mu \rho }\tilde{x}_{\nu }-\eta _{\nu \rho }\tilde{x}_{\mu })+\frac{i\hbar}{\kappa}(a_\nu\tilde{M}_{\mu\rho}-a_\mu\tilde{M}_{\nu\rho}),\label{Yrem1}
\end{equation}
\begin{equation}
[\tilde{M}_{\mu\nu},\tilde{q}_\rho]=i\hbar (\eta _{\mu \rho }\tilde{q}_{\nu }-\eta _{\nu \rho }\tilde{q}_{\mu })+i\hbar\t{\kappa}(b_\nu\tilde{M}_{\mu\rho}-b_\mu\tilde{M}_{\nu\rho}).\label{Yrem22}
\end{equation}
To write the remaining commutation relations we should specify the transformation of the generator $\hat{r}$. Let us consider two cases:
\begin{enumerate}
\item
If we put simply $\t{r}=\h{r}$, this leads to the following modified commutators of Yang algebra
\begin{equation}
[\t{r},\t{x}_\mu]=\frac{i\hbar}{M^2}(\t{q}_\mu-\t{\kappa} b^\rho\t{M}_{\mu\rho}),\qquad [\t{r},\t{q}_\mu]=-\frac{i\hbar}{R^2}(\t{x}_\mu-\frac{1}{\kappa} a^\rho\t{M}_{\mu\rho}),\qquad [\t{r},\t{M}_{\mu\nu}]=0.\label{tri}
\end{equation}
In such a case the internal symmetry and Lorentzian generators are still commuting with each other.
\item
We can introduce the following formula for the modified generator $\h{r}$ (compare with the first term on the RHS of (\ref{Yxp}))
\begin{equation}
\t{r}=\h{r}+\frac{1}{\kappa} a^\rho\t{q}_\rho-\t{\kappa} b^\rho \t{x}_\rho-\frac{\t{\kappa}}{\kappa} a^\rho b^\sigma \t{M}_{\rho\sigma}.\label{2r}
\end{equation}
In this case the commutator (\ref{Yxp}) can be rewritten as follows:
\begin{equation}
[\t{x}_\mu,\t{q}_\nu]=i\hbar\left[\eta_{\mu\nu}\t{r}+\t{\kappa} b_\mu\t{x}_\nu-\frac{1}{\kappa} a_\nu\t{q}_\mu+\frac{\t{\kappa}}{\kappa} a^\rho b_\rho \t{M}_{\mu\nu}\right]\label{Yxp2}.
\end{equation}
We can also calculate all the remaining commutators containing the $\t{r}$ generator
\begin{equation}
[\t{r},\t{x}_\mu]=i\hbar\left[-\frac{1}{\kappa} a_\mu\t{r}-\frac{\t{\kappa}}{\kappa} a^\rho b_\rho\t{x}_\mu+(\frac{1}{M^2}+\frac{a^2}{\kappa^2})\t{q}_\mu\right],\label{d1}
\end{equation}
\begin{equation}
[\t{r},\t{q}_\mu]=i\hbar\left[\t{\kappa} b_\mu\t{r}+\frac{\t{\kappa}}{\kappa}a^\rho b_\rho\t{q}_\mu-(\frac{1}{R^2}+\t{\kappa}^2b^2)\t{x}_\mu\right],\label{d2}
\end{equation}
\begin{equation}
[\t{M}_{\mu\nu}, \t{r}]=i\hbar\left[\frac{1}{\kappa}(a_\mu\t{q}_\nu-a_\nu\t{q}_\mu)-\t{\kappa}(b_\mu\t{x}_\nu-b_\nu\t{x}_\mu)\right]\label{d3}
\end{equation}
and we see from \eqref{d3} that the internal symmetry generator $\h{r}$ and Lorentzian generators do not commute with each other in the new basis.


\end{enumerate}
Further, to obtain the double $\kappa$-deformed Poincar\'e algebra, 
we require the following conditions:
\begin{equation}
\frac{1}{M^2}+\frac{a^2}{\kappa^2}=0,\qquad \frac{1}{R^2}+\t{\kappa}^2b^2=0
\label{cond}
\end{equation}
to remove the terms proportional to $\tilde{M}_{\mu\nu}$ (i.e. the curvature terms)
on the RHS of eqs.\ (\ref{Ygxx}) and (\ref{Ygpp}).

Subsequently, we will consider only two cases\footnote{\tiny{There are 9 choices of double $\kappa$-deformations, which could be considered from mathematical point of view, which correspond to $3\times 3=9$ choices of four-vectors $a_\mu$, $b_\mu$ with their covariant lengths $a^2$, $b^2$ satisfying the three-fold choice $(-1,0,1)$.}} (see e.g. \cite{Blaut}):
\begin{enumerate}
\item{\bf{Doubly time-like $\kappa$ deformations}}\\
If we require $a^2=-\frac{\kappa^2}{M^2}$ then the curvature terms on the RHS of (\ref{Ygxx}) vanish and we get pure $\kappa$-deformation type of commutation for coordinates. Analogously, the condition $b^2=-\frac{1}{R^2\t{\kappa}^2}$ leads to the new $\tilde{\kappa}$ type commutator for four-momenta. The time-like $\kappa$-deformations for coordinates and four-momenta are imposed by the conditions $a^2=-1$ and $b^2=-1$, which due to \eqref{cond} imply the following relations between parameters: $\kappa=M$ and $\t{\kappa}=R^{-1}$. We obtain:
\begin{equation}
[\tilde{x}_\mu,\tilde{x}_\nu]=\frac{i\hbar}{M}(a_\mu\tilde{x}_\nu-a_\nu\tilde{x}_\mu),\qquad
[\tilde{q}_\mu,\tilde{q}_\nu]=\frac{i\hbar}{R}(b_\mu\tilde{q}_\nu-b_\nu\tilde{q}_\mu),\label{Yrp3}
\end{equation}
\begin{equation}
[\tilde{M}_{\mu\nu},\tilde{x}_\rho]=i\hbar (\eta _{\mu \rho }\tilde{x}_{\nu }-\eta _{\nu \rho }\tilde{x}_{\mu })+\frac{i\hbar}{M}(a_\nu\tilde{M}_{\mu\rho}-a_\mu\tilde{M}_{\nu\rho}),
\end{equation}
\begin{equation}
[\tilde{M}_{\mu\nu},\tilde{q}_\rho]=i\hbar (\eta _{\mu \rho }\tilde{q}_{\nu }-\eta _{\nu \rho }\tilde{q}_{\mu })+\frac{i\hbar}{R}(b_\nu\tilde{M}_{\mu\rho}-b_\mu\tilde{M}_{\nu\rho}).\label{YMp3}
\end{equation}
In the case $\t{r}=\hat{r}$, the commutators \eqref{Yrp3}-\eqref{YMp3} are supplemented by the relations
\begin{equation}
[\t{x}_\mu,\t{q}_\nu]=i\eta_{\mu\nu}(\t{r}+\frac{1}{M} a^\rho\t{q}_\rho-\frac{1}{R} b^\rho \t{x}_\rho-\frac{1}{MR} a^\rho b^\sigma \t{M}_{\rho\sigma})+\frac{i}{R} b_\mu\t{x}_\nu-\frac{i}{M} a_\nu\t{q}_\mu+\frac{i}{MR} a^\rho b_\rho \t{M}_{\mu\nu},\label{Yxp3}
\end{equation}
\begin{equation}
[\t{r},\t{x}_\mu]=\frac{i\hbar}{M^2}(\t{q}_\mu-\frac{1}{R}b^\rho\t{M}_{\mu\rho}),\qquad [\t{r},\t{q}_\mu]=-\frac{i\hbar}{R^2}(\t{x}_\mu-\frac{1}{M} a^\rho\t{M}_{\mu\rho}),\qquad [\t{r},\t{M}_{\mu\nu}]=0.
\end{equation}
If we consider $\t{r}$ generator defined by formula (\ref{2r}), the commutators (\ref{Yrp3})-(\ref{YMp3}) should be supplemented by the following ones:
\begin{equation}
[\t{x}_\mu,\t{q}_\nu]=i\hbar\left[\eta_{\mu\nu}\t{r}+\frac{1}{R} b_\mu\t{x}_\nu-\frac{1}{M} a_\nu\t{q}_\mu+\frac{1}{MR} a^\rho b_\rho \t{M}_{\mu\nu}\right],\label{equiv}
\end{equation}
\begin{equation}
[\t{r},\t{x}_\mu]=i\hbar\left[-\frac{1}{M} a_\mu\t{r}-\frac{1}{MR} a^\rho b_\rho\t{x}_\mu\right],\qquad
[\t{r},\t{q}_\mu]=i\hbar\left[\frac{1}{R} b_\mu\t{r}+\frac{1}{MR}a^\rho b_\rho\t{q}_\mu\right],
\end{equation}
\begin{equation}
[\t{M}_{\mu\nu}, \t{r}]=i\hbar\left[\frac{1}{M}(a_\mu\t{q}_\nu-a_\nu\t{q}_\mu)-\frac{1}{R}(b_\mu\t{x}_\nu-b_\nu\t{x}_\mu)\right].
\end{equation}
\item{\bf{Doubly light-cone $\kappa$ deformations}}\\
One can obtain light-cone $\kappa$-deformation for space-time coordinates by imposing $a^2=0$. In such a case we have to take the additional limit $M\rightarrow \infty$ to remove the curvature terms from (\ref{Ygxx}). Analogously (see (\ref{Ygpp})), for the four-momenta we should impose the conditions $b^2=0$ and $R\rightarrow \infty$. We obtain the pair of relations (cf. \eqref{xxk}, \eqref{qqk}):
\begin{equation}
[\tilde{x}_\mu,\tilde{x}_\nu]=\frac{i\hbar}{\kappa}(a_\mu\tilde{x}_\nu-a_\nu\tilde{x}_\mu),\qquad
[\tilde{q}_\mu,\tilde{q}_\nu]=i\hbar\t{\kappa}(b_\mu\tilde{q}_\nu-b_\nu\tilde{q}_\mu).\label{xw2}
\end{equation}
When $\tilde{r}=\hat{r}$, the formulas (\ref{tri}) take the form
\begin{equation}
[\t{r},\t{x}_\mu]= [\t{r},\t{q}_\mu]= [\t{r},\t{M}_{\mu\nu}]=0\label{tri2}
\end{equation}
and because the commutators (\ref{Yxp})-(\ref{Yrem22}) are not $M,R$ -dependent they remain unchanged.
Moreover, from \eqref{tri2} it follows that
$\tilde{r}=\hat{r}\sim \mathrm{I}$ (i.e., it is proportional to the identity operator) and the algebra presented here will correspond to the algebra presented in the next section (see Sec.\ 4, doubly light-cone $\kappa$ deformations).

If we consider generator  $\t{r}$ defined by formula (\ref{2r}), the commutators (\ref{xw2}) are supplemented by the following ones:
\begin{equation}
[\t{r},\t{x}_\mu]=i\hbar\left[-\frac{1}{\kappa} a_\mu\t{r}-\frac{\t{\kappa}}{\kappa} a^\rho b_\rho\t{x}_\mu\right],
\qquad
[\t{r},\t{q}_\mu]=i\hbar\left[\t{\kappa} b_\mu\t{r}+\frac{\t{\kappa}}{\kappa}a^\rho b_\rho\t{q}_\mu\right],
\end{equation}
and we get the formulae (\ref{Yrem1}), (\ref{Yrem22}), (\ref{Yxp2}), (\ref{d3}).
\end{enumerate}

\section{From TSR to new doubly $\kappa$ deformed phase space}
Another way to obtain the doubly $\kappa$ deformed phase spaces is to use a similar approach as the one presented in the previous Sec. 3.3 but now in the case of Triply Special Relativity (TSR) algebra, which was firstly formulated in 2004 \cite{TSR} by the following set of commutation relations:
\begin{equation}
\lbrack \hat{x}_{\mu },\hat{x}_{\nu }]=\frac{i\hbar}{M^2}\hat{M}_{\mu
\nu },  \qquad    \lbrack \hat{p}_{\mu },\hat{p}_{\nu }]=\frac{i\hbar}{R^2}\hat{M}_{\mu
\nu },\qquad M, R\in\mathbb{R}\label{TSR1}
\end{equation}
\begin{equation}
\lbrack \hat{M}_{\mu \nu },\hat{x}_{\rho }]=i\hbar (\eta _{\mu \rho }\hat{x}%
_{\nu }-\eta _{\nu \rho }\hat{x}_{\mu }), \qquad
\lbrack \hat{M}_{\mu \nu },\hat{p}_{\rho }]=i\hbar(\eta _{\mu \rho }\hat{p}%
_{\nu }-\eta _{\nu \rho }\hat{p}_{\mu }), \label{snyderMp}
\end{equation}
\begin{equation}
\lbrack \hat{M}_{\mu \nu },\hat{M}_{\rho \tau }]=i\hbar(\eta _{\mu \rho
}\hat{M}_{\nu \tau }-\eta _{\mu \tau }\hat{M}_{\nu \rho }+\eta _{\nu \tau }\hat{M}_{\mu \rho
}-\eta _{\nu \rho }\hat{M}_{\mu \tau }), \label{snyderMM}
\end{equation}
\begin{equation}
[\hat{x}_\mu,\hat{p}_\nu]=i\hbar\h{g}_{\mu\nu},\qquad \h{g}_{\mu\nu}=\eta_{\mu\nu}+\frac{1}{R^2}\hat{x}_\mu\hat{x}_\nu+\frac{1}{M^2}\hat{p}_\mu\hat{p}_\nu+\frac{1}{MR}(\hat{x}_\mu\hat{p}_\nu+\hat{p}_\mu\hat{x}_\nu-\hat{M}_{\mu\nu}).\label{pxx}
\end{equation}
Formulas \eqref{TSR1}-(\ref{pxx}) satisfy Jacobi identities. We point out that \eqref{pxx} is not unique, and there are infinitely many relations which would satisfy the required Jacobi identities. Note that the subalgebras spanned by the pairs $(\h{M},\h{x})$ and $(\h{M}, \h{p} )$ are just the standard de Sitter algebras.
\\
The generators $\hat{g}=\{\hat{x}_\mu, \hat{p}_\mu,\hat{M}_{\mu\nu}\}$ are Hermitian, i.e. $\hat{g}=\hat{g}^\dagger$. This leads to the following condition
\begin{equation}
\h{g}_{\mu\nu}=\h{g}_{\mu\nu}^\dagger.
\end{equation}
After some straightforward manipulations, we obtain that:
\begin{equation}
g_{\mu\nu}-g_{\mu\nu}^\dagger=\frac{i\hbar}{MR}(\frac{2}{MR}\hat{M}_{\mu\nu}+\h{g}_{\mu\nu}-\h{g}_{\nu\mu}).
\end{equation}
We can compute the expression $\h{g}_{\mu\nu}-\h{g}_{\nu\mu}$ appearing above, by using (\ref{pxx}) and we get \cite{genTSR2}:
\begin{equation}
(\h{g}_{\mu\nu}-\h{g}_{\nu\mu})=-\frac{2}{MR}\hat{M}_{\mu\nu}.\label{sym}
\end{equation}
It means that Hermitian conditions for $\hat{g}$ and  $\h{g}_{\mu\nu}$ are consistent with TSR algebra without any additional assumptions.

Now, we can change the basis of the TSR algebra (\ref{TSR1})-(\ref{pxx}) using linear transformations \eqref{YT}, what leads to the following modified commutators for coordinates and four-momenta
\begin{equation}
[\tilde{x}_\mu,\tilde{x}_\nu]=i\hbar(\frac{1}{M^2}+\frac{a^2}{\kappa^2})\tilde{M}_{\mu\nu}+\frac{i\hbar}{\kappa}(a_\mu\tilde{x}_\nu-a_\nu\tilde{x}_\mu),\label{gxx}
\end{equation}
and
\begin{equation}
[\tilde{p}_\mu,\tilde{p}_\nu]=i\hbar(\frac{1}{R^2}+\t{\kappa}^2 b^2)\tilde{M}_{\mu\nu}+i\hbar\t{\kappa}(b_\mu\tilde{p}_\nu-b_\nu\tilde{p}_\mu),\label{gpp}
\end{equation}
where $a^2=a^\mu a_\mu$ and $b^2=b^\mu b_\mu$, as before.
The remaining commutators of TSR algebra in the new basis described by the formulae \eqref{YT} have the following form:
\begin{equation}
[\tilde{M}_{\mu\nu},\tilde{x}_\rho]=i\hbar (\eta _{\mu \rho }\tilde{x}_{\nu }-\eta _{\nu \rho }\tilde{x}_{\mu })+\frac{i\hbar}{\kappa}(a_\nu\tilde{M}_{\mu\rho}-a_\mu\tilde{M}_{\nu\rho}),\label{rem1}
\end{equation}
\begin{equation}
[\tilde{M}_{\mu\nu},\tilde{p}_\rho]=i\hbar (\eta _{\mu \rho }\tilde{p}_{\nu }-\eta _{\nu \rho }\tilde{p}_{\mu })+i\hbar\t{\kappa}(b_\nu\tilde{M}_{\mu\rho}-b_\mu\tilde{M}_{\nu\rho}),
\end{equation}
\begin{equation}
[\tilde{x}_\mu,\tilde{p}_\nu]=i\hbar\tilde{g}_{\mu\nu},\label{pft}
\end{equation}
\begin{eqnarray}
\tilde{g}_{\mu\nu}=\h{g}_{\mu\nu}(\t{g})+\eta_{\mu\nu}(\frac{1}{\kappa} a^\rho\tilde{p}_\rho-\t{\kappa} b^\rho\tilde{x}_\rho+\frac{\t{\kappa}}{\kappa} a^\rho b^\sigma\tilde{M}_{\sigma\rho})+\t{\kappa} b_\mu\tilde{x}_\nu-\frac{1}{\kappa} a_\nu\tilde{p}_\mu+\frac{\t{\kappa}}{\kappa} ab\tilde{M}_{\mu\nu},\label{tg}
\end{eqnarray}
where we used the shortcut notation for:
\begin{eqnarray}
\h{g}_{\mu\nu}(\t{g})&=&\t{h}_{\mu\nu}+(\frac{1}{\kappa R}a^\rho+\frac{\t{\kappa}}{M}b^\rho)(\frac{1}{\kappa R}a^\sigma+\frac{\t{\kappa}}{M}b^\sigma)\tilde{M}_{\mu\rho}\tilde{M}_{\nu\sigma}\\
&-&(\frac{1}{R}\tilde{x}_\mu+\frac{1}{M}\tilde{p}_\mu)(\frac{1}{\kappa R}a^\rho+\frac{\t{\kappa}}{M}b^\rho)\tilde{M}_{\nu\rho}-(\frac{1}{\kappa R}a^\rho+\frac{\t{\kappa}}{M}b^\rho)\tilde{M}_{\mu\rho}(\frac{1}{R}\tilde{x}_\nu+\frac{1}{M} \tilde{p}_\nu),\nonumber
\end{eqnarray}
and
\begin{equation}
\t{h}_{\mu\nu}=\eta_{\mu\nu}+\frac{1}{R^2}\t{x}_\mu\t{x}_\nu+\frac{1}{M^2}\t{p}_\mu\t{p}_\nu+\frac{1}{MR}(\t{x}_\mu\t{p}_\nu+\t{p}_\mu\t{x}_\nu-\t{M}_{\mu\nu}).\label{tg2}
\end{equation}
One can obtain double $\kappa$-deformed phase space by removing the terms proportional to $\tilde{M}_{\mu\nu}$
on the RHS of (\ref{gxx}) and (\ref{gpp}).

Following our discussion in Sec. 3.3 we will focus only on the two types of double $\kappa$-deformations:
\begin{enumerate}
\item{\bf{Doubly time-like $\kappa$ deformations}}\\
Similarly as in Sec. 3.3 (point 1) to obtain the $\kappa$-deformed commutator \eqref{xxk} for coordinates we must require $a^2=-\frac{\kappa^2}{M^2}$ on the RHS of (\ref{gxx}). Analogously, the $\kappa$-type commutator \eqref{qqk} for four-momenta requires setting $b^2=-\frac{1}{R^2\t{\kappa}^2}$. While $a^2=-1$ and $b^2=-1$ give the time-like (or standard) $\kappa$-deformations for coordinates and four-momenta. This leads to the following relationships between parameters: $\kappa=M$ and $\t{\kappa}=R^{-1}$. This way we obtain the known formulae (see \eqref{xxk}, \eqref{qqk}):
\begin{equation}
[\tilde{x}_\mu,\tilde{x}_\nu]=\frac{i\hbar}{M}(a_\mu\tilde{x}_\nu-a_\nu\tilde{x}_\mu),\qquad
[\tilde{p}_\mu,\tilde{p}_\nu]=\frac{i\hbar}{R}(b_\mu\tilde{p}_\nu-b_\nu\tilde{p}_\mu).\label{rp}
\end{equation}
By imposing additional conditions on four-vectors $a_\mu$ and $b_\mu$, we are able to simplify the formulas (\ref{tg})-(\ref{tg2}).  The simplifications obtained by  removing Lorentz generator $\tilde{M}_{\mu\nu}$ from the right hand side of the commutator (\ref{pft}) can be obtained by putting $a_\mu=-b_\mu$. In this case we can supplement the commutators (\ref{rp}) by
\begin{equation}
[\tilde{M}_{\mu\nu},\tilde{x}_\rho]=i\hbar (\eta _{\mu \rho }\tilde{x}_{\nu }-\eta _{\nu \rho }\tilde{x}_{\mu })+\frac{i\hbar}{M}(a_\nu\tilde{M}_{\mu\rho}-a_\mu\tilde{M}_{\nu\rho}),
\end{equation}
\begin{equation}
[\tilde{M}_{\mu\nu},\tilde{p}_\rho]=i\hbar (\eta _{\mu \rho }\tilde{p}_{\nu }-\eta _{\nu \rho }\tilde{p}_{\mu })+\frac{i\hbar}{R}(b_\nu\tilde{M}_{\mu\rho}-b_\mu\tilde{M}_{\nu\rho}),
\end{equation}
\begin{equation}
[\tilde{x}_\mu,\tilde{p}_\nu]=i\hbar\tilde{g}_{\mu\nu},
\end{equation}
where
\begin{eqnarray}
\tilde{g}_{\mu\nu}&=&\eta_{\mu\nu}(1+\frac{1}{M} a^\rho\tilde{p}_\rho+\frac{1}{R} a^\rho\tilde{x}_\rho)-\frac{1}{R} a_\mu\tilde{x}_\nu-\frac{1}{M} a_\nu\tilde{p}_\mu\cr&&+\frac{1}{R^2}\tilde{x}_\mu\tilde{x}_\nu+\frac{1}{M^2}\tilde{p}_\mu\tilde{p}_\nu+\frac{1}{MR}(\tilde{x}_\mu\tilde{p}_\nu+\tilde{p}_\mu\tilde{x}_\nu).
\end{eqnarray}

\item{\bf{Doubly light-cone $\kappa$ deformations}}\\
Similarly, as in Sec. 3.3 (point 2) to obtain the light-cone $\kappa$-deformation for space-time coordinates we impose $a^2=0$. In such a case, one should take the additional limit $M\rightarrow \infty$ in order to remove the curvature terms proportional to $M_{\mu\nu}$ from (\ref{gxx}). Analogously for four-momenta (see (\ref{gpp})) we have the conditions $b^2=0$ and $R\rightarrow \infty$. In this last case the relations \eqref{xxk}, \eqref{qqk}
are supplemented by formulae (\ref{rem1})-(\ref{tg}), with the condition $\h{g}_{\mu\nu}(\t{g})=\eta_{\mu\nu}.$
If we additionally put $b_\mu=\lambda a_\mu,  \lambda\in\mathbb{R}$, one can remove Lorentz generator $\tilde{M}_{\mu\nu}$ from the right hand side of the commutator (\ref{pft}).
\end{enumerate}

\section{Final remarks}
In this paper we propose new models of doubly $\kappa$-deformed Poincar\'e algebras with two independent $\kappa$-deformations for NC quantum space-time \eqref{xxk} and NC quantum four-momenta coordinates \eqref{qqk}, described by two independent deformation parameters $\kappa$ and $\tilde{\kappa}$. In standard approach, one obtains the
$\kappa$-deformations of space-time coordinates sector by employing the Hopf duality of the $\kappa$-deformed Poincar\'e group and $\kappa$-deformed Poincar\'e algebra. This leads to the description of quantum phase spaces by the Heisenberg double construction \cite{Heis1}-\cite{Heis6}.  
Here, the new doubly $\kappa$-deformed quantum phase spaces are resulting from two various approaches.
We have discussed the following:\\
i) The contraction procedure of double $\kappa$-deformations of Yang model with built in $\hat{o}(2)$ internal symmetry generator $\hat{r}$, what permits to describe algebraically the models with Lie algebraic $\hat{o}(1,5)$ structure. Then by exploring the Lie algebra structure $\hat{o}(1,5;g)$ with the constant parameter-dependent symmetric metric $g$, we have obtained the doubly $\kappa$-deformed quantum phase space by properly fixing some of the parameters  of the metric (namely $g_{44}=g_{55}=0$).
 Additionally, we have also proposed a change of basis directly from the Yang model in order to obtain the  doubly $\kappa$-deformed quantum phase space described algebraically as doubly 
 $\kappa$-deformed Poincar\'e algebras \eqref{xxk}, \eqref{qqk}.\\
ii) As another approach we have proposed a change of basis in the doubly $\kappa$-deformed non-linear TSR model \cite{TSR} which can be obtained from Yang model by expressing bi-linearly the fifteenth generator $\hat{r}$ in terms of the remaining fourteen physical generators $\hat{x}_\mu, \hat{p}_\mu,\hat{M}_{\mu\nu}$.\\
As special cases we discuss doubly time-like and doubly light-cone $\kappa$-deformed quantum phase spaces.

It should be mentioned that the original TSR model \cite{TSR} has been recently generalized \cite{20}, \cite{genTSR2},\cite{genTSR1},\cite{genTSR3} and at present these generalizations are under our consideration. Moreover, the quantum phase spaces with the presence of noncommutativity in both space-time coordinates and four-momenta sectors have been recently considered in the context of Generalized Extended Uncertainty Principle (GEUP) and the analogue of the Liouville theorem \cite{Pachol2024}. The canonical commutation relations of quantum phase spaces which involve both NC space-time coordinates and NC four-momenta, generally will lead to various types of GEUPs (see e.g. \cite{Wagner}, but also \cite{Ghosh:2024eza}) and one can expect that the modifications to the Heisenberg uncertainty relation may result in new thermodynamical properties of statistical systems (see e.g. \cite{Pachol2024}, \cite{PW1}, \cite{PW2}).

\section*{Acknowledgements}
%
J. Lukierski, A. Pacho{\l } and M. Woronowicz acknowledge the support of the
Polish NCN grant 2022/45/B/ST2/01067. S. Mignemi recognizes the support of
Gruppo Nazionale di Fisica Matematica. A. Pacho{\l } acknowledges COST
Action CaLISTA CA21109.

\section*{Data availability}
No data was used for the research described in the article.

\end{document}